\documentclass[aps,prl,showpacs,twocolumn,superscriptaddress]{revtex4}
\usepackage{graphicx}
\usepackage{bm}
\begin{document}

\title{Exact Analysis of Soliton Dynamics in Spinor Bose-Einstein Condensates}
\date{\today}

\author{Jun'ichi Ieda}
\email{ieda@monet.phys.s.u-tokyo.ac.jp}
\affiliation{Department of Physics, Graduate School of Science, University of Tokyo, Bunkyo-ku, Tokyo 113-0033, Japan}
\author{Takahiko Miyakawa}
\email{tmiyakawa@optics.arizona.edu}
\affiliation{Department of Physics, Graduate School of Science,
University of Tokyo, Bunkyo-ku, Tokyo 113-0033, Japan}
\affiliation{Optical Sciences Center, University of Arizona, Tucson, AZ 85721, USA}
\author{Miki Wadati}
\affiliation{Department of Physics, Graduate School of Science,
University of Tokyo, Bunkyo-ku, Tokyo 113-0033, Japan}

\begin{abstract}
We propose an integrable model of a multicomponent spinor Bose-Einstein
condensate in one dimension, which allows an exact description of the 
dynamics of bright solitons with spin degrees of freedom.
We consider specifically an atomic condensate in the $F=1$ hyperfine state
confined by an optical dipole trap.
When the mean-field interaction is attractive ($c_0 < 0$) 
and the spin-exchange interaction of a spinor condensate is 
ferromagnetic ($c_2 < 0$),
we prove that the system possesses a completely integrable point
leading to the existence of multiple bright solitons.
By applying results from the inverse scattering method,
we analyze a collision law for two-soliton solutions and
find that the dynamics can be explained in terms of
the spin precession.
\end{abstract}



\maketitle

In 2002, two groups~\cite{SolRice,SolEns} reported matter-wave
solitons of an atomic Bose-Einstein condensate (BEC).
They prepared BECs of ${}^7\textrm{Li}$ atoms in a small region of an elongated
optical dipole trap, which is an analog of a waveguide for microwaves.
After tuning the strength of interaction between the atoms to a sufficiently
large negative value, they set the condensate free along the waveguide.
The solitary wave packets were formed and propagated
in the guide nondispersively.
It is well known that the Gross-Pitaevskii (GP) equation with attractive
interactions in a one-dimensional (1D) space,
which is also called the self-focusing nonlinear
Schr\"odinger (NLS) equation, has bright soliton solutions~\cite{refsoliton}.
Therefore, they concluded that the dynamics of the system is actually
1D so that the matter-wave solitons can be
observed.

Matter-wave solitons in a new field of atom optics~\cite{Meystre}
are expected to be useful for applications in atom laser,
atom interferometry, and coherent atom transport.
Moreover, it could contribute to the realization of quantum
information processing or computation.
When one explores these future applications, atomic BECs have
another advantage.
That is, atoms have many internal degrees of freedom liberated under
an optical trap~\cite{MIT}, giving rise to a multiplicity of signals.
The properties of BECs with spin degrees of freedom were
investigated by many researchers~\cite{MIT,Ho,SBECs,spindynamics}.

In this Letter, we combine these two fascinating properties,
matter-wave soliton and internal degrees of freedom.
We consider BECs of alkali atoms in the $F=1$ hyperfine state, 
such as ${}^7\textrm{Li}$, ${}^{87}\textrm{Rb}$, and ${}^{23}\textrm{Na}$,
confined in the 1D space by purely optical means.
Under no external magnetic fields, their three internal states
$m_F=1,0,-1,$ where $m_F$ is the magnetic quantum number, are degenerate.
The dynamics of the spinor condensates is described by the 
multicomponent GP equations within the mean-field approximation.
Those coupled equations have nontrivial nonlinear terms reflecting
the $SU(2)$ symmetry of the spins.

When the mean-field interaction is attractive and the spin-exchange
interaction is ferromagnetic, we show that this system possesses a 
completely integrable point.
By considering a reduction of results from
the inverse scattering method, we present for the first time
the exact multiple bright soliton solutions for the system with
the spin-exchange interaction.
The spin-exchange interaction, which is absent in the systems
of Ref.~\cite{SolRice,SolEns} because of
frozen spin degrees of freedom under additional magnetic fields,
gives rise to spin mixing within condensates~\cite{spindynamics}
during soliton-soliton collisions.
We analyze the collision law for two-soliton solutions
and find that the soliton dynamics can be explained in terms of
the spin precession.

The assembly of atoms in the $F=1$ state is characterized by
a vectorial order parameter:
${\bm \Phi}(x,t)\equiv[\Phi_1(x,t),\Phi_0(x,t),\Phi_{-1}(x,t)]^T$
with the components subject to the hyperfine spin space.
The normalization is imposed as
$\int dx\,{\bm \Phi}(x,t) ^{\dagger}\cdot{\bm \Phi}(x,t)=N_T$,
where  $N_T$ is the total number of atoms.
Here we assume that the system is one dimensional: 
the trap is elongated in the $x$ direction such that the transverse
spatial degrees of freedom are factorized from the longitudinal
and all the hyperfine states are in the transverse ground state.
This quasi-one-dimensional regime is achievable~\cite{Carr}.
The interaction between atoms in the $F=1$ hyperfine state is
given by
$V(x_1-x_2)=\delta(x_1-x_2)(\bar{c}_0+\bar{c}_2{\bm F}_1\cdot{\bm F}_2)$
, where ${\bm F}_j$ are the angular momentum of two atoms \cite{Ho}.
In this expression,
$\bar{c}_0=(\bar{g}_0+2\bar{g}_2)/3$, $\bar{c}_2=(\bar{g}_2- \bar{g}_0)/3$,
with the effective 1D couplings,~\cite{CIR}
\begin{equation}
\bar{g}_f=\frac{4\hbar^2a_f}{m a_\perp^2}\frac{1}{(1-C a_f/a_\perp)},\nonumber
\end{equation}
where $a_f$ are the $s$-wave scattering lengths in the total hyperfine
spin $f$ channel, $a_\perp$ is the size of the transverse ground states,
$m$ is the atomic mass, and $C=-\zeta(1/2)=1.4603\cdots$.
Then, the Gross-Pitaevskii energy functional is expressed as
\begin{eqnarray}
\label{GPenergy}
E_{\scriptsize \textrm{GP}}&=& \int  d x \,\left(\frac{\hbar^2}{2m}\partial_x
\Phi^*_\alpha \partial_x\Phi_\alpha+
\frac{\bar{c}_0}{2}\Phi^*_\alpha
\Phi^*_{\alpha^\prime} \Phi_{\alpha^\prime}
\Phi_\alpha \right.\nonumber\\ 
&& \left.{}+\frac{\bar{c}_2}{2}\Phi^*_\alpha
\Phi^*_{\alpha^\prime}{\bm f}_{\alpha \beta}^T \cdot
{\bm f}_{\alpha^\prime \beta^\prime}\Phi_{\beta^\prime}
\Phi_{\beta}\right),
\end{eqnarray}
where repeated subscripts $\{\alpha,\beta,\alpha',\beta'=1,0,-1\}$
should be summed up and ${\bm f}= (f^x,\,f^y,\,f^z )^T$ with
$f^i$ being $3\times 3$ spin-1 matrices.

The time evolution of spinor condensate wave function ${\bm \Phi}(x,t)$
can be derived from the variational principle:
$\textrm{i}\hbar\partial_t \Phi_\alpha(x,t)=
\delta E_{\scriptsize \textrm{GP}}/\delta \Phi^*_\alpha(x,t)$.
Substituting Eq.~(\ref{GPenergy}) into this,
we obtain a set of equations:
\begin{eqnarray}
\textrm{i}\hbar\partial_t\Phi_{\pm 1}
&=& -\frac{\hbar^2}{2m}\partial^2_x\Phi_{\pm 1}+
(\bar{c}_0+\bar{c}_2)(|\Phi_{\pm 1}|^2+|\Phi_{0}|^2)\Phi_{\pm 1}\nonumber\\
&&{}+(\bar{c}_0-\bar{c}_2)|\Phi_{\mp 1}|^2\Phi_{\pm 1}
+\bar{c}_2\Phi^*_{\mp 1}\Phi^2_0,\nonumber\\
\textrm{i}\hbar\partial_t\Phi_0
&=&-\frac{\hbar^2}{2m}\partial^2_x\Phi_0+\bar{c}_0|\Phi_0|^2\Phi_0+
(\bar{c}_0+\bar{c}_2)(|\Phi_{1}|^2\nonumber\\
&&{}+|\Phi_{-1}|^2)\Phi_0+2\bar{c}_2\Phi^*_0\Phi_{1}\Phi_{-1}.
\label{TDeq}
\end{eqnarray}

In this Letter, we consider the system with the coupling constants
$\bar{c}_0=\bar{c}_2\equiv-c<0$, equivalently
$2\bar{g}_0=-\bar{g}_2>0$.
The effective interactions between atoms in a BEC have been
tuned with a Feshbach resonance~\cite{Mag}.
In spinor BECs, however, we should extend this to alternative techniques
such as an optically induced Feshbach resonance~\cite{Opt} or a confinement
induced resonance~\cite{CIR}, which do not affect the rotational symmetry
of the internal spin states.
In the latter, the above condition is surely obtained by setting
$a_\perp=3Ca_0a_2/(2a_0+a_2)$ in Eq.~(1) when $a_0>a_2>0$ or $a_2>0>a_0$.
Recently, such strong transversal confinement has been realized in a 2D
optical lattice where $a_\perp\sim$ tens of $nm$~\cite{2Dopt latt}.

Introducing the dimensionless form,
$\bm{\Phi}\to (\phi_1, \sqrt{2} \phi_0,$ $\phi_{-1})^T$,
where time and length are measured 
in units of $\bar{t}=\hbar a_\perp/c$ and $\bar{x}=\hbar\sqrt{a_\perp/2mc}$,
respectively, we can rewrite Eqs.~(\ref{TDeq}) as a
$2 \times 2$ matrix version of the NLS equation:
\begin{equation}
\label{NLS}
\textrm{i}\partial_t Q+ \partial^2_x Q+2QQ^\dagger Q={\mathit O},
\quad
Q=\left(\begin{array}{cc}
\phi_1 & \phi_0 \\
\phi_0 & \phi_{-1}
\end{array}
\right).
\end{equation}
Since the matrix NLS equation~(\ref{NLS}) is integrable~\cite{Tsuchida1},
the dynamical problems of this system can be solved exactly.
The embedding (\ref{NLS}) and its first application
to an atomic system with the spin-exchange interaction
are the main idea of this Letter.

We remark that a different choice of $Q$ \cite{Tsuchida1}
gives rise to coupled NLS equations known as
the Manakov model~\cite{Manakov}
which is widely used to describe the interaction among
the modes in nonlinear optics.

The general $N$-soliton solution of Eq.~(\ref{NLS})
is obtained through a reduction of a formula derived by
the inverse scattering method (ISM) in \cite{Tsuchida1} as
\begin{eqnarray}
\label{N soliton}
Q(x,t)=(\,\underbrace{I\,\cdots I}_N\,)S^{-1}
\left(
\begin{array}{c}
\Pi_1 e^{\chi_1} \\
\vdots \\
\Pi_N e^{\chi_N}
\end{array}
\right),
\end{eqnarray}
where $I$ is the $2\times 2$ unit matrix and
the $2N\times 2N$ matrix $S$ is given by
\begin{eqnarray*}
S_{ij}=\delta_{ij}I+\sum_{l=1}^N
       \frac{\Pi_i\cdot\Pi_l^\dagger}{(k_i+k_l^*)(k_j+k_l^*)}
       e^{\chi_i+\chi_l^*},
\end{eqnarray*}
($1\le i,j \le N$).
Here we have introduced the following:
\begin{eqnarray*}
\Pi_j&=& \left(\begin{array}{cc}
\beta_j&\alpha_j\\
\alpha_j&\gamma_j
\end{array}\right),\,\,
2|\alpha_j|^2+|\beta_j|^2+|\gamma_j|^2=1,
\nonumber\\
\chi_j&\equiv&\chi_j(x,t)=k_jx+\textrm{i} k_j^2t-\epsilon_j.
\end{eqnarray*}
The $2\times2$ matrices $\Pi_j$ normalized to unity in the sense of the 
square norm must take the same form as $Q$ from their definition.
We call them ``polarization matrices", which determine both 
the populations of the three components $\{1,\,0,\,-1\}$ within each soliton
and the relative phases between them.
The complex constant $k_j$ denotes discrete eigenvalue of $j$th soliton,
which determines a bound state by
the potential $Q$ in the context of ISM~\cite{Tsuchida1}.
$\epsilon_j$ is a real constant which can be used to
tune the initial displacement of soliton.

The equation (\ref{NLS}) is a completely integrable system
in the sense that the initial value problems can be solved
via ISM.
The existence of the $\mathbf r$ matrix for this system 
guarantees the existence of an infinite number of conservation
laws~\cite{Tsuchida1}
which restrict the dynamics of the system in an essential way.
Here we show explicit forms of some conserved quantities:
\textit{number}, $N_T= \int d x \,n(x,t)$,
$n(x,t)={\bm\Phi}^\dagger\cdot{\bm\Phi}=\textrm{tr}\{Q^\dagger Q\}$;
\textit{spin}, ${\mathbf F}_T= \int d x \,{\mathbf f}(x,t)$,
${\mathbf f}(x,t)={\bm\Phi}^\dagger\!\cdot\!
\bm{f}\!\cdot\!\bm{\Phi}
=\textrm{tr}\{Q^{\dagger} {\bm \sigma} Q\}$, 
(${\bm \sigma}$: Pauli matrices);
\textit{momentum},
$P_T= \int d x \,p(x,t)$,
$p(x,t)= -\textrm{i}\hbar{\bm\Phi}^\dagger\cdot\partial_x{\bm\Phi}
= -\textrm{i}\hbar \,\textrm{tr}\{Q^\dagger Q_x\}$;
\textit{energy},
$E_T= \int d x \,e(x,t)$,
$e(x,t)= (\hbar^2/2m)\partial_x{\bm\Phi}^\dagger\cdot\partial_x{\bm\Phi}
-{}c\left(n^2+{\mathbf f}^2 \right)/2\
= c\,\textrm{tr}\{Q^\dagger_{x}Q_{x}-Q^\dagger QQ^\dagger Q\}$.

If we set $N=1$ in the formula (\ref{N soliton}),
we obtain the one-soliton solution:
\begin{eqnarray}
\label{OSS}
Q=2k_R\frac{\Pi\,e^{-(\chi_R+\rho/2)}+\left(\sigma^y\Pi^\dagger\sigma^y\right)e^{\chi_R+\rho/2}\det\Pi}
{e^{-(2\chi_R+\rho)}+1
+e^{2\chi_R+\rho} |\det\Pi|^2}e^{{\scriptsize \textrm{i}} \chi_I},
\end{eqnarray}
where $e^{\rho/2}\equiv (2k_R)^{-1}$,
and the subscripts \textit{R} and \textit{I}
denote real and imaginary parts, respectively.
We set $k_R>0$ without loss of generality.

In Eq.~(\ref{OSS}),
we can make out the significance
of each parameter or coordinate as follows:
$k_R$, amplitude of soliton;
$ 2k_I$, velocity of soliton's envelope;
$\chi_R$, coordinate for observing soliton's envelope;
$\chi_I$, coordinate for observing soliton's carrier waves.
We use the term ``amplitude" in the sense of the peak(s) height
of soliton's envelope.
The actual amplitude should be represented as $k_R$ multiplied by a factor
from 1 to $\sqrt{2}$ which is determined by the type of polarization matrices.
Note that the motion of soliton depends on both $x$ and $t$ via
variables $\chi_R$ and $\chi_I$,
which elucidates the meaning of $2k_I$ as a velocity.

Because of the spin conservation, the one-soliton solution
can be classified by the spin states.
We show that only two spin states are allowable,
i.e., $|\mathbf{F}_T|=N_T$ for $\det \Pi=0$, and
$|\mathbf{F}_T|=0$ for $\det\Pi\ne 0$.

\textit{Ferromagnetic state}.---
Under the condition $\det\Pi=0$ ($\alpha^2=\beta\gamma$),
Eq.~(\ref{OSS}) becomes a simple form:
\begin{eqnarray*}
Q=k_R\textrm{sech}(\chi_R+\rho/2)\Pi e^{{\scriptsize \textrm{i}}\chi_I}.
\end{eqnarray*}
Now all of the $m_F=0,\,\pm1$ components share the same wave function.
Their distribution in the internal states reflects the elements of
the polarization matrix $\Pi$ directly. One can clearly see
the meaning of each parameter listed above.
The number of particles is calculated as $N_T=2k_R$.
The spin of this soliton becomes
\begin{eqnarray}
\label{ferrospin}
\mathbf{F}_T=2k_R\textrm{tr}\{\Pi^\dagger {\bm \sigma}\Pi\}
,\quad|\mathbf{F}_T|=N_T.
\end{eqnarray}
Thus, this type of soliton belongs to the ferromagnetic state.
The momentum and the energy of
the ferromagnetic state are
$P_T^f=N_T\hbar k_I$, 
$E_T^f=N_T c ( k_I^2-k_R^2/3)$.

\textit{Polar state}.---
In the case of $\det\Pi\ne 0$, a local spin density has one node,
i.e., $\mathbf{f}(x_0,t)=0$ at a point
$x_0= 2k_It+[\ln(4k_R^2/|\det\Pi|)
+2\epsilon]/2k_R$, for each moment of $t$.
Setting $x^\prime=x-x_0$ and $A^{-1}=2|\det\Pi|$, we obtain
\begin{eqnarray*}
\mathbf{f}(x^\prime)=-\frac{4k_R^2 A \textrm{sinh}(2k_Rx^\prime)}
{\big[A+\cosh(2k_Rx^\prime)\big]^2}
\textrm{tr}\{\Pi^\dagger {\bm \sigma}\Pi\}.
\end{eqnarray*}
Since each component of the local spin density is an odd function
of $x^\prime$, its average value becomes zero, i.e.,
$\mathbf{F}_T=(0,0,0)^T$.
This implies that this type of soliton, on the average,
belongs to the polar state~\cite{Ho}. 
Note that the relation $N_T=4k_R$ is
different from that of the ferromagnetic state.
The momentum and the energy are given by
$P_T^p=N_T\hbar k_I$,
$E_T^p=N_T c \left( k_I^2-k_R^2/3\right)$,
respectively.
The energy difference between the ferromagnetic state and the polar state
with the same number of particles $N_T$ is
$E_T^f-E_T^p=-N_T^3 c/16<0$,
which is a natural consequence of the ferromagnetic interaction,
i.e., $\bar{c}_2<0$.

The two-soliton solution can be obtained by setting $N=2$ in
Eq.~(\ref{N soliton}). Since the derivation is straightforward but
lengthy, we give an explicit formula of the general two-soliton solution
in a separate paper~\cite{IMWfull}, and here, focus on the two-soliton
solution in the 
energetically favorable ferromagnetic state ($\det \Pi_1=\det\Pi_2=0$),
computing the asymptotic forms as $t\to \mp \infty$, which define
a collision law of two solitons in the spinor model.
For simplicity, we confine the spectral parameters to regions
$k_{1R}>0$, $k_{2R}<0$, $k_{1I}<0$, $k_{2I}>0$,
which correspond to a head-on collision.
Under these conditions, we calculate the asymptotic forms in the final state
($t\to\infty$) from those in the initial state ($t\to -\infty$).
In the asymptotic forms, we can consider each soliton separately.
Thus, the initial state is given by a sum of two solitons as
$Q\simeq Q_1^{\scriptsize \textrm{in}}+Q_2^{\scriptsize \textrm{in}}$,
where
$Q_j^{\scriptsize \textrm{in}}=
k_{jR}\textrm{sech}(\chi_{jR}+\rho_j/2)\Pi_j
e^{{\scriptsize \textrm{i}}\chi_{jI}}$.
And for the final state,
$Q\simeq Q_1^{\scriptsize \textrm{fin}}+Q_2^{\scriptsize \textrm{fin}}$,
where
$
Q_j^{\scriptsize \textrm{fin}}=
k_{jR}\textrm{sech}\left(\chi_{jR}+\rho_j/2+s\right)
{\tilde \Pi}_j e^{{\scriptsize \textrm{i}}\chi_{jI}}$.
Here we have introduced
the phase shift $s=\ln [1-(4k_{1R}k_{2R}/|k_1+k_2^*|^2)
|\textrm{tr}(\Pi_1\Pi_2^\dagger)|]$
and
the polarization matrices $\tilde{\Pi}_j$ in the final state.
Each polarization matrix $\Pi_j$ of the ferromagnetic state
can be expressed by three real variables $\tau_j,\theta_j,\varphi_j$~\cite{Ho}, as
\begin{eqnarray*}
\Pi_{j}= e^{{\scriptsize \textrm{i}}\tau_j}\left(\begin{array}{cc}
\cos^2\frac{\theta_j}{2} e^{-{\scriptsize \textrm{i}}\varphi_j} &
\cos\frac{\theta_j}{2}\sin\frac{\theta_j}{2} \\
\cos\frac{\theta_j}{2}\sin\frac{\theta_j}{2} &
\sin^2\frac{\theta_j}{2} e^{{\scriptsize \textrm{i}}\varphi_j} 
\end{array}\right).
\end{eqnarray*}
In this expression, we have the following collision law:
\begin{eqnarray*}
\Pi_{j}= e^{{\scriptsize \textrm{i}}\tau_j}\mathbf{u}_j \cdot \mathbf{u}^T_j,\,\,
{\tilde\Pi}_{j}=
 e^{-s+{\scriptsize \textrm{i}}\tau_j}{\tilde\mathbf{u}}_j\cdot{\tilde\mathbf{u}}^T_j,
\end{eqnarray*}
where with $(j, l)=(1,2)$, $(2,1)$,
\begin{eqnarray*}
\mathbf{u}_{j}=\left(\begin{array}{cc}
\cos{\frac{\theta_j}{2}} e^{-{\scriptsize \textrm{i}}\frac{\varphi_j}{2}} \\
\sin{\frac{\theta_j}{2}} e^{{\scriptsize \textrm{i}}\frac{\varphi_j}{2}} \end{array}\right)\!,\,\,
{\tilde\mathbf{u}}_j=\mathbf{u}_j-\frac{k_l+k_l^*}{k_j+k_l^*}
(\mathbf{u}^\dagger_l \cdot \mathbf{u}_j ) \mathbf{u}_l.
\end{eqnarray*}

\begin{figure}[b]
  \begin{center}
    \includegraphics[width=8cm,height=8cm]{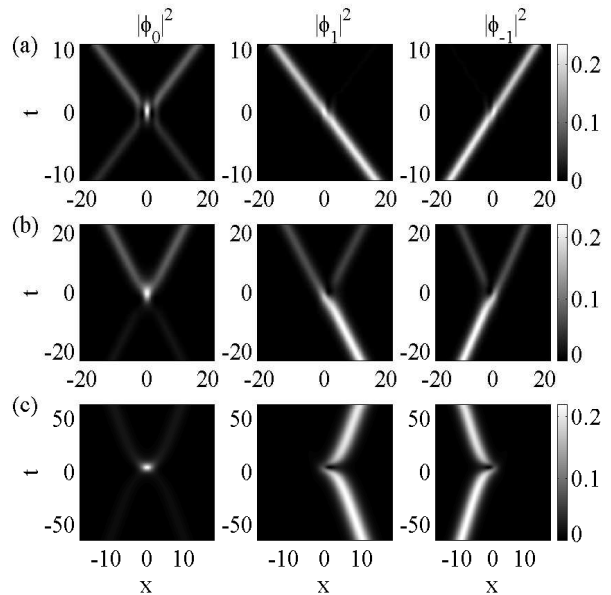}
  \end{center}
  \vspace*{-5mm}
  \caption{Time evolution of $|\phi_0|^2$ (left column),
  $|\phi_{1}|^2$ (middle column), $|\phi_{-1}|^2$ (right column)
  for (a) $k_I=0.75$, (b) $0.25$, and (c) $0.05$,
  with $k_R=0.5$, $\alpha_{1,2}=4/17$,
  $\beta_1=\gamma_2=16/17$, and $\gamma_1=\beta_2=1/17$.}
  \label{lfig1}
\end{figure}

Since each envelope is located around $x\simeq2k_{jI}t$,
soliton 1 and soliton 2 are initially isolated at $x\to\pm\infty$
and then travel to the opposite direction at a velocity
of $2k_{1I}$ and $2k_{2I}$, respectively.
After a head-on collision, they pass through without changing their
amplitudes and velocities and arrive at $x\to\mp\infty$
in the final state.
The collision induces rotations of their polarizations 
in addition to the usual phase shifts.
The collision laws for other cases, two-soliton of polar-ferromagnetic
(${\det}\Pi_1\neq 0, \det\Pi_2=0$) and polar-polar
($\det\Pi_1\ne 0, \det\Pi_2\ne 0$),
can be obtained in the same way \cite{IMWfull}.

Figure~\ref{lfig1} shows the time evolution of the density profiles
for $\{1,0,-1\}$ components in three different velocities:
(a) $-k_{1I}=k_{2I}\equiv k_I=0.75$, (b) $k_I=0.25$, and (c) $k_I=0.05$, with
$k_{1R}=-k_{2R}\equiv k_R=0.5$, $\alpha_{1,2}=4/17$, $\beta_1=\gamma_2=16/17$, and
$\gamma_1=\beta_2=1/17$.
In the initial state, soliton 1 (left mover)
consists mostly of the $m_F=1$ component
and, on the contrary, soliton 2 (right mover) almost lies in $m_F=-1$.
A fast collision, Fig.~\ref{lfig1}(a), makes the solitons almost
transparent to each other.
As $k_I$ decreases, the residence time inside the collisional region
increases, and the mixing among the components occurs in each
outgoing soliton. In Fig.~\ref{lfig1}(c), the components are switched
between the solitons after their collision.
As seen in Fig.~\ref{lfig1}(b) clearly, the number of each component
can vary not only in each soliton but also in the total during the collision
in consequence of the spin-exchange interaction.
This contrasts to the Manakov system~\cite{Manakov}, where the total number of each
component is conserved.

We can gain a better understanding of the two-soliton collision by
recasting it in terms of the spin dynamics.
The total spin conservation restricts
the motion of the spin of each soliton on a circumference
around the total spin axis.
Since a spin of the ferromagnetic soliton is given by Eq.~(\ref{ferrospin}),
that of the $j$th soliton in the initial state is
$\mathbf{F}_{j}=2|k_{jR}|(\sin{\theta_j}\cos{\varphi_j},
\sin{\theta_j}\sin{\varphi_j},\cos{\theta_j})^T$.
When we set $|k_{1R}|=|k_{2R}|\equiv N_T/4$,
the final state spins $\tilde\mathbf{F}_j$ are obtained through
$\mathbf{F}_{1,2}$ by 
$\tilde\mathbf{F}_j=\cos^2(\omega/2)\mathbf{F}_j+
\sin^2(\omega/2)\mathbf{F}_l+\sin{\omega} 
(\mathbf{F}_j \times \mathbf{F}_l)/|\mathbf{F}_T|$,
where $\mathbf{F}_T=\mathbf{F}_1+\mathbf{F}_2$, and $\omega$ is a rotation angle
of the spin precession.
The rotation angle $\omega$ is determined only by the ratio $k_I/k_R$
and the magnitude of the normalized total spin ${\mathcal F}\equiv
|\mathbf{F}_T/N_T|$ as
$\cos{(\omega/2)}= e^{-s/2}$ with
$e^{s}=1+(k_R/k_I)^2\, {\mathcal F}^2$.

\begin{figure}
  \begin{center}
    \includegraphics[width=7cm,height=4.5cm]{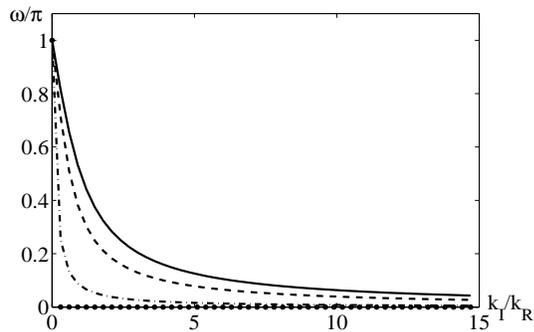}
  \end{center}
  \vspace*{-5mm}
  \caption{$\omega$ versus $k_I/k_R$ for ${\mathcal F}=1$ (solid line),
    $0.5$ (dashed line), $0.0157$ (dash-dot line),
    and $0$ (dotted line).}
  \label{lfig2}
\end{figure}

Figure~\ref{lfig2} shows $\omega$ as a function of the ratio
$k_I/k_R$ for different values of the normalized total spin: ${\mathcal F}=1$,
0.5, 0.0157, and 0.
In consistency with Fig.~\ref{lfig1}, it exhibits that $\omega$ becomes
larger as $k_I/k_R$ decreases.
The large (small) total spin makes the spin of each soliton rotate a lot (bit).
When $\mathbf{F}_T$ is zero, corresponding to the case of antiparallel
spin collision, the spin precession cannot occur as shown by the
dotted line in Fig.~\ref{lfig2}.

In this Letter, we have introduced the integrable model 
which describes the dynamics of $F=1$ spinor
BECs in one dimension.
Utilizing the inverse scattering method,
we have obtained the multiple soliton solutions.
One-soliton solutions are classified into two distinct spin
states: ferromagnetic, $|{\mathbf F}_T|=N_T$ and
polar, $|{\mathbf F}_T|=0$.
We have also shown the collision law for
the two solitons of the ferromagnetic state and identified their
collision with the spin precession dynamics around the total spin.
We believe these properties should be observed in experiment and
lead to a variety of applications such as coherent atom transport
and quantum information.
\begin{acknowledgments}
J.~I. thanks T.~Tsuchida for many useful discussions. 
\end{acknowledgments}

\end{document}